\documentclass[aps,prd,twocolumn,preprintnumbers,floatfix,nofootinbib,superscriptaddress]{revtex4-2}

\pdfoutput=1
\usepackage[T1]{fontenc}
\usepackage[utf8]{inputenc}
\usepackage{graphicx}
\usepackage{xcolor}
\definecolor{revisiongreen}{RGB}{0,110,60}
\usepackage{amsmath,amssymb,amsfonts}
\usepackage{booktabs}
\usepackage{multirow}
\usepackage{array}
\usepackage{slashed}
\usepackage{bm}
\usepackage{hyperref}
\usepackage{tikz}
\usepackage{tikz-feynman}
\usepackage{adjustbox}

\tikzfeynmanset{compat=1.1.0}

\hypersetup{
	colorlinks=true,
	linkcolor=blue,
	citecolor=blue,
	urlcolor=blue
}

\providecommand{\openone}{\leavevmode\hbox{\small1\kern-3.8pt\normalsize1}}

\newcommand{\met}{\slashed{E}_T}

\newcommand{\MG}{\textsc{MadGraph5\_aMC@NLO}}
\newcommand{\PYTHIA}{\textsc{Pythia}~8.30}
\newcommand{\DELPHES}{\textsc{Delphes}~3.5.0}
\newcommand{\MA}{\textsc{MadAnalysis5}}
\newcommand{\XGB}{\textsc{XGBoost}}

\begin{document}
	
	\preprint{IFJPAN-IV-2026-12}
	
	\title{Sensitivity to Single Vector-Like Bottom Quark Production at 3 TeV CLIC}
	
	\author{R. Benbrik}
	\email{r.benbrik@uca.ac.ma}
	\affiliation{Polydisciplinary Faculty, Laboratory of Physics, Energy, Environment, and Applications, Cadi Ayyad University, Sidi Bouzid, B.P. 4162, Safi, Morocco}
	
	\author{M. Berrouj}
	\email{mbark.berrouj@ced.uca.ma}
	\affiliation{Polydisciplinary Faculty, Laboratory of Physics, Energy, Environment, and Applications, Cadi Ayyad University, Sidi Bouzid, B.P. 4162, Safi, Morocco}
	
	\author{M. Boukidi}
	\email{mohammed.boukidi@ifj.edu.pl}
	\affiliation{Institute of Nuclear Physics, Polish Academy of Sciences, ul. Radzikowskiego 152, Cracow, 31-342, Poland}
	
	\author{M. Ech-chaouy}
	\email{m.echchaouy.ced@uca.ma}
	\affiliation{Polydisciplinary Faculty, Laboratory of Physics, Energy, Environment, and Applications, Cadi Ayyad University, Sidi Bouzid, B.P. 4162, Safi, Morocco}
	
	\author{K. Kahime}
	\email{Kahimek@gmail.com}
	\affiliation{Laboratoire Interdisciplinaire de Recherche en Environnement, Management, Energie et Tourisme (LIREMET), ESTE, Cadi Ayyad University, B.P. 383, Essaouira, Morocco}
	
	\author{K. Salime}
	\email{k.salime.ced@uca.ma}
	\affiliation{Polydisciplinary Faculty, Laboratory of Physics, Energy, Environment, and Applications, Cadi Ayyad University, Sidi Bouzid, B.P. 4162, Safi, Morocco}
	
\begin{abstract}
		We study the sensitivity of the Compact Linear Collider at $\sqrt{s}=3~\mathrm{TeV}$ to the single electroweak production of a vector-like bottom quark in a type-II Two-Higgs-Doublet Model extended by vector-like quarks. We consider the process $e^+e^-\to B\bar b+\bar B b$, followed by $B\to\phi b$ with $\phi=H,A$ and $\phi\to t\bar t$. We focus on the semileptonic final state with one charged lepton, missing transverse momentum, four $b$-tagged jets, and two light jets. This channel directly probes the interplay between the extended quark and Higgs sectors. After applying theoretical constraints, electroweak precision constraints, Higgs-sector bounds, and current collider limits, we perform detector-level cut-based and multivariate analyses. The cut-based analysis remains below discovery sensitivity at $5~\mathrm{ab}^{-1}$, while a multilayer perceptron and an XGBoost classifier provide substantially improved signal-background separation. For viable scenarios with $m_B\simeq1.3$--$1.5~\mathrm{TeV}$, the expected significance remains above $5\sigma$ even for a $15\%$ uncertainty on the background normalization.
\end{abstract}
	
	\maketitle
	
\section{Introduction}

	The discovery of a Higgs boson by the ATLAS and CMS Collaborations~\cite{ATLAS:2012yve,CMS:2012qbp} established the existence of a scalar particle associated with electroweak symmetry breaking. Measurements of its properties are so far consistent with the Standard Model (SM), but they do not determine whether the observed state is the only scalar degree of freedom. The one-doublet Higgs sector is the minimal realization of electroweak symmetry breaking rather than a structure enforced by a fundamental principle. Establishing whether the scalar sector is minimal, or instead contains additional neutral and charged states, therefore remains an important goal of the collider program.
	
	Extended scalar sectors arise naturally in several well-motivated theories beyond the SM. The Minimal Supersymmetric Standard Model (MSSM), for example, requires two Higgs doublets and predicts additional neutral and charged Higgs bosons~\cite{Djouadi:2005gj}. Composite-Higgs constructions may also contain extra scalar resonances, and suitable symmetry-breaking patterns can give rise to a second Higgs doublet~\cite{Bellazzini:2014yua,Mrazek:2011iu}. A simple and widely used low-energy framework for studying such nonminimal scalar dynamics is the Two-Higgs-Doublet Model (2HDM)~\cite{Gunion:1992hs,Branco:2011iw}. After electroweak symmetry breaking, its physical spectrum contains two CP-even neutral scalars, $h$ and $H$, one CP-odd scalar, $A$, and a charged Higgs-boson pair, $H^\pm$.
	
	The additional Higgs states may be produced directly or through the decays of heavier particles. The latter possibility is especially relevant when the scalar sector is accompanied by new fermionic degrees of freedom. Cascade decays of heavy fermions can provide a complementary production mechanism for non-SM Higgs bosons and lead to collider signatures that are absent in the SM~\cite{Gopalakrishna:2015wwa,Benbrik:2019zdp,Arhrib:2024nbj,Benbrik:2023xlo,Benbrik:2025nfw}. Such processes probe the scalar and fermion sectors simultaneously and may access parameter regions that are difficult to test through direct heavy-Higgs production alone.
	
	Vector-like quarks (VLQs) are among the simplest and most extensively studied extensions of the SM fermion sector~\cite{Aguilar-Saavedra:2009xmz,Okada:2012gy,Buchkremer:2013bha}. In contrast to chiral SM quarks, their left- and right-handed components transform in the same way under the SM gauge group, allowing gauge-invariant mass terms independently of electroweak symmetry breaking. VLQs arise naturally in theories with extra dimensions~\cite{Chang:1999nh,Gherghetta:2000qt,Contino:2003ve}, Little Higgs models~\cite{Arkani-Hamed:2002iiv,Schmaltz:2002wx,Chang:2003vs,Han:2003wu}, and composite-Higgs scenarios~\cite{Agashe:2004rs,Bellazzini:2014yua,Contino:2006qr,Lodone:2008yy,Matsedonskyi:2012ym}. Depending on their electroweak representation, they may appear as singlets, doublets, or triplets containing states with electric charges $+2/3$, $-1/3$, $+5/3$, and $-4/3$, conventionally denoted by $T$, $B$, $X$, and $Y$, respectively.
	
	Most current collider searches for VLQs target the standard decay modes
	\begin{equation}
		T\to Wb, Zt, ht,
		\qquad
		B\to Wt, Zb, hb,
	\end{equation}
	while the exotic states predominantly decay through $X\to Wt$ and $Y\to Wb$~\cite{CMS:2026iyo,ATLAS:2024zlo,ATLAS:2023ixh,CMS:2024bni,CMS:2022fck,ATLAS:2024xdc,CMS:2021mku,ATLAS:2022tla,ATLAS:2024gyc,ATLAS:2024kgp}. The corresponding limits are commonly interpreted under the assumption that decays into the SM gauge bosons and the SM-like Higgs boson saturate the total VLQ width. Although this assumption is appropriate in minimal scenarios, it need not hold in the presence of an extended scalar sector.
	
	If additional neutral or charged Higgs bosons are kinematically accessible, new decay modes such as
	\begin{equation}
		T\to Ht, At, H^+b,
		\qquad
		B\to Hb, Ab, H^-t
	\end{equation}
	become possible~\cite{Bhardwaj:2022nko,Bardhan:2022sif,Benbrik:2026zyu,Benbrik:2025kvz,Benbrik:2022kpo}. These channels can compete with, or in parts of the parameter space dominate over, the conventional decays into $W$, $Z$, and $h$. The resulting branching-ratio pattern may therefore differ substantially from the one assumed in standard searches, reducing the direct applicability of the corresponding experimental limits. At the same time, the nonstandard channels lead to final states containing additional Higgs bosons, top quarks, and several heavy-flavor jets, which are not fully covered by conventional VLQ analyses.
	
	The 2HDM extended by VLQs provides a simple framework in which to study this interplay. The additional Higgs bosons modify the available VLQ decay channels, while VLQ decays offer a complementary mechanism for producing the heavy scalar states~\cite{Gopalakrishna:2015wwa,Benbrik:2019zdp,Benbrik:2024hsf,Arhrib:2024nbj,Benbrik:2023xlo,Benbrik:2025nfw,Benbrik:2022kpo,Benbrik:2024bxt,Arhrib:2024mbq,Arhrib:2024dou,Arhrib:2024tzm,Arhrib:2016rlj,Abouabid:2023mbu,Angelescu:2015uiz,Ghosh:2023xhs,Dermisek:2019vkc,Cingiloglu:2023ylm,Banerjee:2016wls,Dermisek:2020gbr,Dermisek:2021zjd,Benbrik:2026zjv,Benbrik:2026zgr}. Dedicated searches for these cascade decays are therefore necessary for a more complete exploration of both the VLQ and extended-Higgs sectors.
	
	Future high-energy lepton colliders provide a clean environment for studying such signatures. The Compact Linear Collider (CLIC) is foreseen to operate at center-of-mass energies of $380~\mathrm{GeV}$, $1.5~\mathrm{TeV}$, and $3~\mathrm{TeV}$~\cite{CLICDetector:2013tfe,CLIC:2018fvx}. At the highest-energy stage, CLIC can probe new states at the TeV scale while benefiting from a well-defined initial state, reduced QCD backgrounds, and cleaner event reconstruction than at a hadron collider. These features are particularly useful for final states containing several heavy-flavor jets and top quarks, for which the backgrounds and combinatorial ambiguities can otherwise be substantial.
	
	In this work, we study the single production of a vector-like bottom quark in the $(T,B)$ doublet realization of the type-II 2HDM with VLQs at CLIC with $\sqrt{s}=3~\mathrm{TeV}$. We consider
	\begin{equation}
		e^+e^-\to B\bar b+\bar B b,
	\end{equation}
	followed by the nonstandard decay $B\to\phi b$, with $\phi=H,A$, and the subsequent decay $\phi\to t\bar t$. We focus on the semileptonic channel in which one top quark decays leptonically and the other hadronically, leading to a final state with one charged lepton, missing transverse momentum, four $b$-tagged jets, and two light jets. This process probes a heavy VLQ and an additional neutral Higgs boson within the same event and therefore provides a direct test of the connection between the extended fermion and scalar sectors.
	
	We first perform a cut-based analysis using the most discriminating kinematic observables. Because the signal and background distributions overlap in several variables, we then investigate whether multivariate methods can exploit their correlations more effectively. We employ a multilayer perceptron and a gradient-boosted decision tree implemented with XGBoost and compare their performance with the cut-based strategy. This allows us to assess the sensitivity of CLIC to nonstandard VLQ decays that are not directly targeted by searches based on the conventional $W$, $Z$, and SM-like Higgs-boson channels.
	
	The paper is organized as follows. Section~\ref{sec:Framework} presents the theoretical framework and the relevant vector-like-bottom-quark interactions. Section~\ref{sec:Constraints} summarizes the theoretical, electroweak, Higgs-sector, and collider constraints imposed on the parameter space. Section~\ref{sec:cutbased} describes the benchmark selection, simulation setup, and cut-based analysis. Section~\ref{sec:ML} presents the multivariate analysis and the resulting sensitivity. Our conclusions are given in Sec.~\ref{sec:Conclusions}.

	\section{Theoretical framework}
	\label{sec:Framework}
	
	This section outlines the theoretical setup of the 2HDM-II extended by VLQs. The scalar sector is taken to be a CP-conserving 2HDM with two $SU(2)_L$ doublets, $\Phi_1$ and $\Phi_2$, both carrying hypercharge $Y=+1$. To control tree-level flavor-changing neutral currents, a discrete $\mathbb{Z}_2$ symmetry, $\Phi_1\to-\Phi_1$, is usually imposed and softly broken in the scalar potential. The most general gauge-invariant scalar potential compatible with these assumptions is~\cite{Branco:2011iw,Gunion:1989we}
	\begin{align}
		V(\Phi_1,\Phi_2) &= m_{11}^2\Phi_1^\dagger\Phi_1 + m_{22}^2\Phi_2^\dagger\Phi_2 - m_{12}^2(\Phi_1^\dagger\Phi_2 + \Phi_2^\dagger\Phi_1) \notag \\
		&\quad + \frac{\lambda_1}{2}(\Phi_1^\dagger\Phi_1)^2 + \frac{\lambda_2}{2}(\Phi_2^\dagger\Phi_2)^2 + \lambda_3(\Phi_1^\dagger\Phi_1)(\Phi_2^\dagger\Phi_2) \notag \\
		&\quad + \lambda_4(\Phi_1^\dagger\Phi_2)(\Phi_2^\dagger\Phi_1) + \frac{\lambda_5}{2}\left[(\Phi_1^\dagger\Phi_2)^2+(\Phi_2^\dagger\Phi_1)^2\right],
		\label{thdmV}
	\end{align}
	where all parameters are taken to be real. The two complex doublets may be rotated into the Higgs basis $(H_1,H_2)$, in which only one doublet acquires a vacuum expectation value (VEV). After electroweak symmetry breaking and after applying the minimization conditions of the potential, the physical scalar sector can be described by seven independent parameters. A convenient choice is given by the four physical Higgs masses, $(m_h,m_H,m_A,m_{H^\pm})$, the ratio of VEVs $\tan\beta=v_2/v_1$, the CP-even mixing parameter $\cos(\beta-\alpha)$, and the soft $\mathbb{Z}_2$-breaking parameter $m_{12}^2$.
	
	VLQs are exotic fermions whose left- and right-handed components transform identically under the SM gauge group $SU(3)_c\times SU(2)_L\times U(1)_Y$. Depending on their electroweak quantum numbers, these states may be embedded in different multiplet representations:
	\begin{align}
		& T^0_{L,R},\quad B^0_{L,R} && \text{(singlets)}, \notag \\
		& (X\,T^0)_{L,R},\quad (T^0\,B^0)_{L,R} && \text{(doublets)}, \notag \\
		& (X\,T^0\,B^0)_{L,R},\quad (T^0\,B^0\,Y)_{L,R} && \text{(triplets)}.
	\end{align}
	Here the superscript $0$ denotes weak eigenstates, while the corresponding fields without the superscript denote mass eigenstates. The field $B^0$ carries electric charge $Q=-1/3$.
	
	In this work, we focus on a vector-like bottom quark belonging to a $(T,B)$ doublet. The presence of $B^0_{L,R}$ modifies the down-type quark sector and yields four mass eigenstates, $d$, $s$, $b$, and $B$. The mixing is assumed to occur dominantly with the third generation, as mixings with the first two generations are strongly constrained by flavor observables and precision measurements, including LEP constraints on $R_b$~\cite{Aguilar-Saavedra:2002phh}. The mixing between $b^0$ and $B^0$ is parametrized as
	\begin{eqnarray}
		\left( \begin{array}{c}
			b_{L,R} \\
			B_{L,R}
		\end{array} \right)
		=
		\left( \begin{array}{cc}
			\cos\theta_{L,R}^d & -\sin\theta_{L,R}^d e^{i\phi_d} \\
			\sin\theta_{L,R}^d e^{-i\phi_d} & \cos\theta_{L,R}^d
		\end{array} \right)
		\left( \begin{array}{c}
			b^0_{L,R} \\
			B^0_{L,R}
		\end{array} \right),
		\label{ec:mixd}
	\end{eqnarray}
	where $\theta_{L,R}^d$ are the left- and right-handed mixing angles, and $\phi_d$ is a CP-violating phase. In the present analysis this phase is neglected.
	
	The Yukawa sector in the Higgs basis contains the terms
	\begin{equation}
		-\mathcal{L}_Y \supset y^u\bar Q_L^0\widetilde H_2 u_R^0 + y^d\bar Q_L^0 H_1 d_R^0 + M_u^0\bar u_L^0u_R^0 + M_d^0\bar d_L^0d_R^0 + \text{h.c.},
	\end{equation}
	with $u_R^0=(u_R,c_R,t_R,T_R)$ and $d_R^0=(d_R,s_R,b_R,B_R)$. The VLB mass matrix is written as
	\begin{eqnarray}
		\mathcal{L}_{\rm mass} = -\left( \begin{array}{cc}
			\bar b_L^0 & \bar B_L^0
		\end{array} \right)
		\left( \begin{array}{cc}
			y_{33}^d\,v/\sqrt{2} & y_{34}^d\,v/\sqrt{2} \\
			y_{43}^d\,v/\sqrt{2} & M^0
		\end{array} \right)
		\left( \begin{array}{c}
			b_R^0 \\
			B_R^0
		\end{array} \right)+\text{h.c.},
		\label{ec:Lmass}
	\end{eqnarray}
	where $M^0$ is a bare vector-like mass term and $y_{ij}^d$ are Yukawa couplings. The matrix is diagonalized through a bi-unitary transformation,
	\begin{equation}
		U_L^d\mathcal{M}^d(U_R^d)^\dagger=\mathcal{M}_{\rm diag}^d.
		\label{ec:diag}
	\end{equation}
	
	The relation between the mixing angles depends on the electroweak representation of the VLQ multiplet. For the $SU(2)_L$ doublet considered here, the right-handed mixing angle is given by
	\begin{eqnarray}
		\tan 2\theta_R^d &=& \frac{\sqrt{2}\,|y_{43}^d|\,v\,M^0}{(M^0)^2-\frac{1}{2}v^2\left(|y_{33}^d|^2+|y_{43}^d|^2\right)}.
		\label{ec:angle1}
	\end{eqnarray}
	The left- and right-handed mixing angles are further related by
	\begin{eqnarray}
		\tan\theta_L^d &=& \frac{m_b}{m_B}\tan\theta_R^d,
		\label{ec:rel-angle1}
	\end{eqnarray}
	where $m_b$ is the SM bottom-quark mass and $m_B$ is the VLB mass.
	
	The partial decay width of the VLB into a neutral Higgs state $\phi=H,A$ and a bottom quark is
	\begin{align}
		\Gamma(B\to\phi b) &= \frac{g^2}{128\pi}\frac{m_B}{M_W^2}\,
		\lambda^{1/2}(m_B^2,m_b^2,m_\phi^2) \notag \\
		&\quad\times \Bigg[
		\left(|Y^L_{\phi bB}|^2+|Y^R_{\phi bB}|^2\right)(1+r_b^2-r_\phi^2) \notag \\
		&\qquad\qquad \pm 4r_b\,\mathrm{Re}\!\left(Y^L_{\phi bB}Y^{R*}_{\phi bB}\right)
		\Bigg],
		\label{eq:GammaB}
	\end{align}
	where $r_b=m_b/m_B$, $r_\phi=m_\phi/m_B$, and the sign depends on the CP nature of the neutral scalar. In the alignment limit, $\sin(\beta-\alpha)=1$, the relevant couplings are
	\begin{equation}
		Y^L_{\phi bB}=\tan\beta\,s_R^d c_R^d,
		\qquad
		Y^R_{\phi bB}=\frac{m_b}{m_B}Y^L_{\phi bB}.
	\end{equation}
	
	The partial decay width of a neutral Higgs boson $\phi=H,A$ into a top-quark pair is~\cite{Djouadi:2005gj}
	\begin{equation}
		\Gamma(\phi\to t\bar t)=
		N_c\frac{G_Fm_t^2}{4\sqrt{2}\pi}\,
		g_{\phi t\bar t}^2\,m_\phi\,\beta_t^{\,p},
	\end{equation}
	where $\beta_t=\sqrt{1-4m_t^2/m_\phi^2}$, with $p=3$ for a CP-even scalar and $p=1$ for a CP-odd scalar. In the alignment limit of the 2HDM+VLQ setup, the neutral-Higgs couplings to top quarks, normalized to the SM Higgs coupling, are taken as
	\begin{equation}
		g_{H\bar tt}=-\cot\beta\left[1+(c_R^u)^2\right],
		\qquad
		g_{A\bar tt}=\cot\beta\left[1+(c_R^u)^2\right].
	\end{equation}
	
	\subsection{Experimental and theoretical constraints}
	\label{sec:Constraints}
	
	We impose theoretical and experimental requirements on the parameter space in order to ensure consistency with perturbative unitarity, vacuum stability, electroweak precision data, Higgs measurements, and collider limits.
	
	\subsubsection*{Theoretical constraints}
	
	\begin{itemize}
		\item \textbf{Perturbative unitarity:} The $S$-wave amplitudes for scalar-scalar, scalar-gauge, and gauge-gauge scattering must satisfy perturbative unitarity in the high-energy limit~\cite{Kanemura:1993hm}.
		
		\item \textbf{Perturbativity of the scalar sector:} The quartic couplings entering the scalar potential are required to satisfy $|\lambda_i|<8\pi$ for $i=1,\ldots,5$~\cite{Branco:2011iw}, ensuring the validity of the perturbative expansion.
		
		\item \textbf{Vacuum stability:} The scalar potential must be bounded from below in all field directions. This leads to the conditions~\cite{Deshpande:1977rw,Barroso:2013awa}
		\begin{align}
			&\lambda_1>0,\qquad \lambda_2>0,
			\qquad \lambda_3>-\sqrt{\lambda_1\lambda_2}, \notag \\
			&\lambda_3+\lambda_4-|\lambda_5|>-\sqrt{\lambda_1\lambda_2}.
		\end{align}
		
		\item \textbf{Electroweak precision observables:}
		We use the 2026 Particle Data Group global fit to the oblique parameters, with $U=0$~\cite{ParticleDataGroup:2026aaa},
		\begin{align}
			S_0&=0.008\pm0.071, &
			T_0&=0.021\pm0.055, &
			\rho_{ST}&=0.92.
		\end{align}
		For each parameter point, the scalar and VLQ contributions are combined as $S=S_{\rm 2HDM}+S_{\rm VLQ}$ and $T=T_{\rm 2HDM}+T_{\rm VLQ}$. We evaluate
		\begin{equation}
			\chi^2_{ST}=\Delta\mathbf{x}^{\rm T}C^{-1}\Delta\mathbf{x},
			\qquad
			\Delta\mathbf{x}=\begin{pmatrix}S-S_0\\ T-T_0\end{pmatrix},
		\end{equation}
		where $C$ is the covariance matrix constructed from the quoted uncertainties and correlation coefficient. We retain points satisfying $\chi^2_{ST}\leq5.99$, corresponding to the $95\%$ confidence region for two degrees of freedom. The VLQ contributions are computed using the analytic expressions of Ref.~\cite{Arhrib:2024tzm}. The updated constraint is applied point by point to the parameter scan; the three benchmark points in Table~\ref{tab:BPs} remain inside the allowed ellipse. The calculation is implemented in a modified version of \texttt{2HDMC-1.8.0}~\cite{Eriksson:2009ws}, extended to include VLQ effects as described in Refs.~\cite{Benbrik:2022kpo,Abouabid:2023mbu}.
	\end{itemize}
	
	\subsubsection*{Experimental constraints}
	
	\begin{itemize}
		\item \textbf{Searches for additional Higgs bosons:} Constraints on heavy neutral scalars, $H$ and $A$, and charged Higgs bosons, $H^\pm$, are imposed using \texttt{HiggsBounds-6}~\cite{Bechtle:2008jh,Bechtle:2011sb,Bechtle:2013wla,Bechtle:2015pma}, embedded in the \texttt{HiggsTools} framework~\cite{Bahl:2022igd}. This ensures compatibility with exclusion limits from LEP, the Tevatron, and the LHC.
		
		\item \textbf{SM-like Higgs measurements:} Compatibility with the observed scalar state near $125~\mathrm{GeV}$ is tested using \texttt{HiggsSignals-3}~\cite{Bechtle:2020uwn,Bechtle:2020pkv}, also through \texttt{HiggsTools}. We require $\Delta\chi^2\leq 6.18$ at the $95\%$ confidence level over the set of 159 signal-strength measurements.
		
		\item \textbf{$b\to s\gamma$:} In the ordinary 2HDM-II, the radiative transition $b\to s\gamma$ imposes a strong lower bound, $m_{H^\pm}\gtrsim580~\mathrm{GeV}$. In the presence of VLQs, this bound can be relaxed through loop-induced cancellations. For example, in the $(T,B)$ doublet case, viable configurations with $m_{H^\pm}\sim360~\mathrm{GeV}$ can arise depending on the mixing structure~\cite{Benbrik:2022kpo}. In the present analysis, we adopt the conservative requirement $m_{H^\pm}\geq600~\mathrm{GeV}$.
		
		\item \textbf{LHC constraints on VLQs:} Constraints on the vector-like bottom quark from LHC searches are evaluated by comparing the theoretical and observed production cross sections. We retain only points satisfying $\sigma_{\rm theo}/\sigma_{\rm obs}<1$, following the procedure detailed in Ref.~\cite{Benbrik:2024fku}.
	\end{itemize}
	
	\subsection{Production and decay of VLB quarks}
	
\begin{figure}[t]
		\centering
		\begin{tikzpicture}[scale=0.54]
			\begin{feynman}
				\vertex (e_pos) at (-5, 2) {$\mathbf{e^{+}}$};
				\vertex (e_neg) at (-5, -2) {$\mathbf{e^{-}}$};
				\vertex (v1) at (-2, 0);
				\vertex (v2) at (1, 0);
				\vertex (b_direct) at (4, -2.0) {$\mathbf{b}$};
				\vertex (v4) at (4, 2);
				\vertex (phi) at (6, 3);
				\vertex (b_from_B) at (6.5, 1.5) {$\mathbf{b}$};
				\vertex (t) at (8, 4) {$\mathbf{t}$};
				\vertex (tbar) at (8, 2.5) {$\mathbf{\bar{t}}$};
				\diagram* {
					(e_neg) -- [fermion, ultra thick] (v1) -- [fermion, ultra thick] (e_pos),
					(v1) -- [boson, ultra thick, blue, edge label=$\mathbf{Z}$] (v2),
					(v2) -- [fermion, green!60!black, ultra thick, edge label'=$\mathbf{B}$] (v4),
					(v2) -- [fermion, ultra thick] (b_direct),
					(v4) -- [scalar, red, ultra thick, dashed, edge label={\textcolor{red}{$\mathbf{\phi}$}}] (phi),
					(v4) -- [fermion, ultra thick] (b_from_B),
					(phi) -- [fermion, ultra thick] (t),
					(phi) -- [anti fermion, ultra thick] (tbar),
				};
			\end{feynman}
		\end{tikzpicture}
		\caption{Representative Feynman diagram for single vector-like bottom-quark production followed by the cascade decay $B\to b\phi$ and $\phi\to t\bar t$.}
		\label{fig:Bprod}
	\end{figure}
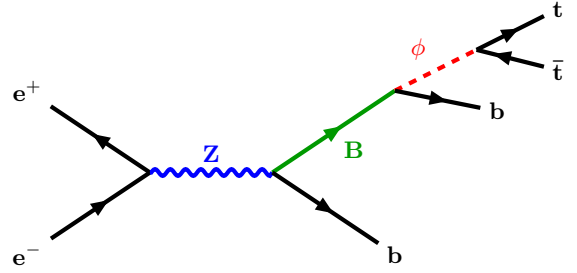

	The single production of a VLB at CLIC proceeds through electroweak interactions and is dominated by an $s$-channel neutral gauge-boson exchange, as illustrated in Fig.~\ref{fig:Bprod}. The produced VLB subsequently decays into a bottom quark and a heavy neutral scalar, $\phi=H,A$.
	
\begin{figure*}[t]
		\centering
		\includegraphics[width=1\textwidth]{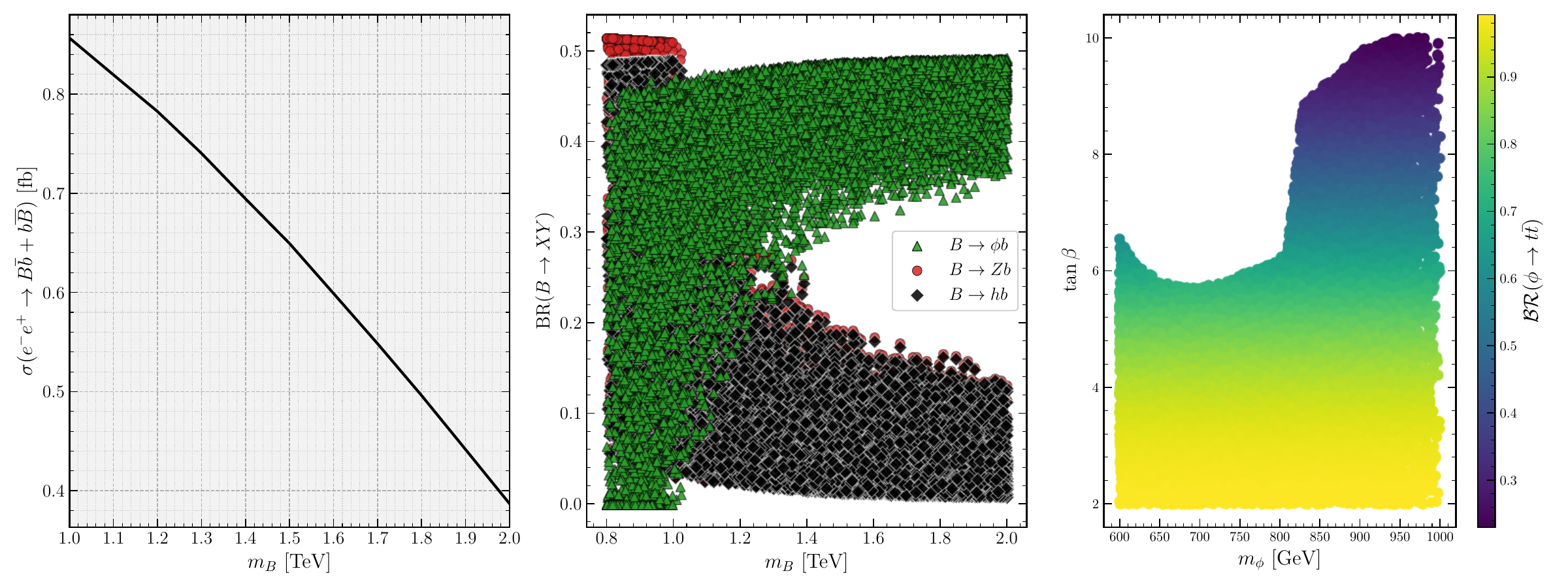}
		\caption{Left: leading-order cross section $\sigma(e^+e^-\to B\bar b+\bar B b)$ as a function of $m_B$ for $m_H=m_A=m_{H^\pm}=600~\mathrm{GeV}$, $\tan\beta=3.5$, $s_R^u=0.01$, and $s_R^d=0.25$ at $\sqrt{s}=3~\mathrm{TeV}$. Middle: $B$-quark branching ratios as functions of $m_B$. Right: ${\rm BR}(\phi\to t\bar t)$ in the $(m_\phi,\tan\beta)$ plane. For the middle and right panels, $m_H=m_A=m_{H^\pm}\in[600,1000]~\mathrm{GeV}$, $m_B\in[0.8,2]~\mathrm{TeV}$, $\tan\beta\in[2,10]$, $s_R^u=0.01$, and $s_R^d=0.25$.}
		\label{fig:xs_br}
	\end{figure*}

	In the left panel of Fig.~\ref{fig:xs_br}, we show the leading-order cross section for $e^+e^-\to B\bar b+\bar B b$ as a function of $m_B$, computed with \MG~(\texttt{MG5})~\cite{Alwall:2014hca}. The cross section decreases from $0.86~\mathrm{fb}$ at $m_B=1~\mathrm{TeV}$ to $0.39~\mathrm{fb}$ at $m_B=2~\mathrm{TeV}$.
	
	The middle panel of Fig.~\ref{fig:xs_br} shows the $B$-quark branching ratios as functions of $m_B$, while the right panel shows ${\rm BR}(\phi\to t\bar t)$ in the $(m_\phi,\tan\beta)$ plane. We scan $m_H=m_A=m_{H^\pm}\in[600,1000]~\mathrm{GeV}$, $m_B\in[0.8,2]~\mathrm{TeV}$, and $\tan\beta\in[2,10]$, with $s_R^u=0.01$ and $s_R^d=0.25$. Since $m_\phi$ and $\tan\beta$ are varied at the same time, the spread at a fixed $m_B$ comes from both phase space and the coupling dependence. Close to the $B\to\phi b$ threshold, the heavy-scalar channels are suppressed and the standard modes $B\to Zb$ and $B\to hb$ can still be important. At larger $m_B$, where more phase space is available, the $\tan\beta$ enhancement of the $B\phi b$ coupling allows ${\rm BR}(B\to Hb)$ and ${\rm BR}(B\to Ab)$ to reach about $48\%$ each. The plot should therefore not be read as showing that the heavy-scalar modes dominate over the whole scan.

	This behavior follows from the couplings. The leading $B\phi b$ coupling scales as $\tan\beta\,s_R^d c_R^d$, while the $BZb$ and $Bhb$ couplings do not receive the same $\tan\beta$ enhancement. For the mixing hierarchy used here, the charged-current modes $B\to Wt$ and $B\to H^-t$ are too small to be visible on the scale of the plot. They are therefore omitted from the legend, but they are still included in the total width. The branching fraction ${\rm BR}(\phi\to t\bar t)$ is largest at low $\tan\beta$, as expected from the $\cot\beta$ dependence of the top-quark Yukawa coupling~\cite{BENBRIK2026117436}.

	\section{Cut-based analysis}
	\label{sec:cutbased}
	
	We select benchmark points satisfying all theoretical and experimental constraints. In particular, we choose $s_d^R\gg s_u^R$ in order to enhance the signal rate, since the branching ratio scales approximately as ${\rm BR}(B\to\phi b)\propto s_d^R c_d^R$. The benchmark points listed in Table~\ref{tab:BPs} also satisfy $|{\rm BR}(B\to Ab)-{\rm BR}(B\to Hb)|\lesssim3\%$ and $|{\rm BR}(H\to t\bar t)-{\rm BR}(A\to t\bar t)|\lesssim3\%$. Consequently, the signal cross sections associated with the two heavy neutral Higgs bosons are nearly identical. A dedicated analysis of either neutral state therefore yields essentially the same sensitivity, and we present the collider study in terms of a generic heavy neutral scalar $\phi$.
	
	The signal process considered in this study is
	\begin{equation}
		e^+e^-\to B\bar b\to \phi b\bar b\to t\bar t b\bar b\to \ell^+ + \met + 4b + 2j.
	\end{equation}
	The dominant SM backgrounds arise from the following processes:
	\begin{itemize}
		\item $e^+e^-\to t\bar tbb$, including $e^+e^-\to t\bar tbbj$;
		\item $e^+e^-\to t\bar tj$, including $e^+e^-\to t\bar tjj$;
		\item $e^+e^-\to tW^-\bar b$ and $e^+e^-\to \bar tW^+b$, with the subsequent decays $t\to bW^+$, $\bar t\to\bar bW^-$, and one $W$ boson decaying leptonically;
		\item $e^+e^-\to W^+W^-V$, with $V=Z,h$, $V\to b\bar b$, and one $W$ boson decaying leptonically while the other decays hadronically.
	\end{itemize}
	
	\begin{table}[t]
		\centering
		\begin{adjustbox}{width=0.78\columnwidth}
			\begin{tabular}{lccc}
				\toprule
				Parameter & BP$_1$ & BP$_2$ & BP$_3$ \\
				\midrule
				$m_h$ [GeV]  & 125.09 & 125.09 & 125.09 \\
				$m_H$ [GeV]  & 745.84 & 780.79 & 664.31 \\
				$m_A$ [GeV]  & 753.52 & 786.08 & 727.33 \\
				$m_{H^\pm}$ [GeV] & 943.37 & 866.79 & 740.56 \\
				$\tan\beta$ & 3.447 & 3.81 & 3.942 \\
				$\sin(\beta-\alpha)$ & 1 & 1 & 1 \\
				$m_T$ [GeV] & 1031.55 & 1384.17 & 1774.33 \\
				$m_B$ [GeV] & 1065.32 & 1429.49 & 1832.43 \\
				$s_L^u$ & 0.0016 & 0.0012 & 0.0009 \\
				$s_L^d$ & 0.0010 & 0.0007 & 0.0005 \\
				$s_R^u$ & 0.01 & 0.01 & 0.01 \\
				$s_R^d$ & 0.25 & 0.25 & 0.25 \\
				\midrule
				\multicolumn{4}{c}{Branching ratios [\%]} \\
				\midrule
				${\cal BR}(B\to Hb)$ & 38.36 & 44.02 & 46.81 \\
				${\cal BR}(B\to Ab)$ & 36.85 & 43.52 & 44.04 \\
				${\cal BR}(H\to t\bar t)$ & 85.83 & 81.16 & 79.58 \\
				${\cal BR}(A\to t\bar t)$ & 88.76 & 84.56 & 83.31 \\
				\bottomrule
			\end{tabular}
		\end{adjustbox}
		\caption{Benchmark points for the 2HDM+$TB$ scenario. Masses are given in GeV and branching ratios in percent.}
		\label{tab:BPs}
	\end{table}
	
	The simulation chain is based on \MG~for the hard-scattering event generation. Parton showering and hadronization are performed with \PYTHIA~\cite{Sjostrand:2014zea}. Detector effects are modeled using \DELPHES~\cite{deFavereau:2013fsa} with a dedicated CLIC detector card. Jets are reconstructed with the Valencia algorithm (VLC)~\cite{Boronat:2014hva,Boronat:2016tgd} in inclusive mode, using a radius parameter $R=0.7$. For $b$-jet identification, we adopt a loose working point corresponding to a $b$-tagging efficiency of $70\%$. The final event selection and cut-flow analysis are performed in the \MA~framework~\cite{Conte:2013mea}.
	
	\begin{figure*}[t]
		\centering
		\includegraphics[width=0.85\textwidth]{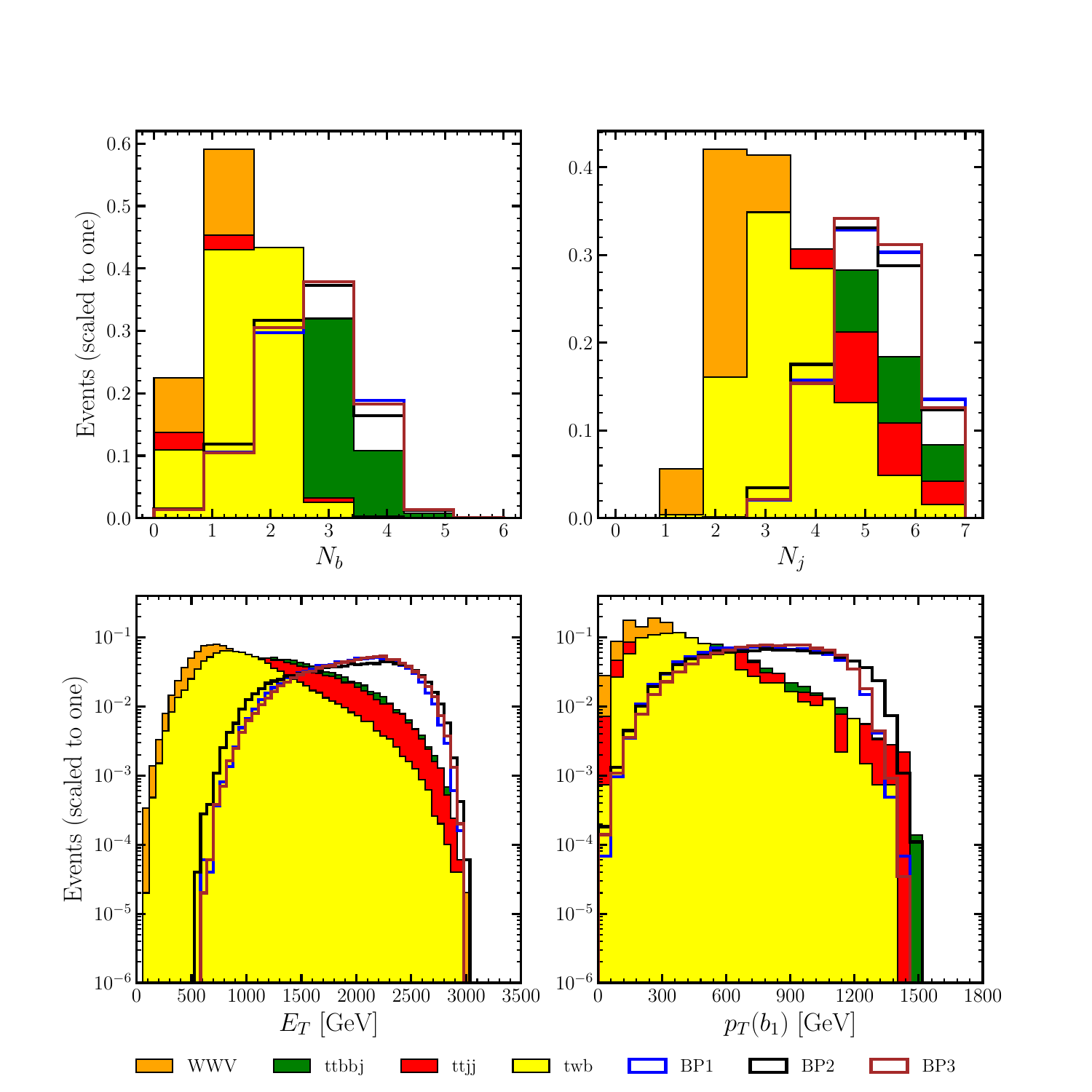}
		\caption{Normalized distributions of representative kinematic observables for the signal and the dominant SM backgrounds at $\sqrt{s}=3~\mathrm{TeV}$.}
		\label{fig:obs}
	\end{figure*}
	
	At parton level, we impose the following baseline cuts on both the signal and background samples:
	\begin{itemize}
		\item $p_T^{\ell}>10~\mathrm{GeV}$ and $p_T^{j,b}>20~\mathrm{GeV}$,
		\item $|\eta_{\ell,b}|<2.5$ and $|\eta_j|<5$,
		\item $\Delta R(x,y)>0.4$ for all $x,y=j,b,\ell$.
	\end{itemize}
	Here $p_T^{\ell,b,j}$ and $\eta^{\ell,b,j}$ denote the transverse momentum and pseudorapidity of leptons, $b$-jets, and light jets, respectively. The angular separation is defined as $\Delta R=\sqrt{\Delta\phi^2+\Delta\eta^2}$.
	
	In Fig.~\ref{fig:obs}, we display the distributions of several discriminating observables for the signal and background samples. The upper-left panel shows the number of $b$-tagged jets, the upper-right panel displays the total number of jets, the lower-left panel shows the total transverse energy $E_T$, and the lower-right panel illustrates the transverse momentum of the leading $b$-jet, $p_T(b_1)$. Guided by these distributions, we define the cut-based selection summarized in Table~\ref{tab:cuts}.
	
	\begin{table}[t]
		\centering
		\begin{tabular}{lc}
			\toprule
			Cut & Definition \\
			\midrule
			Cut 1 & $N(b)\geq3$, $N(j)\geq5$ \\
			Cut 2 & $E_T>1600~\mathrm{GeV}$ \\
			Cut 3 & $p_T(b_1)>500~\mathrm{GeV}$, $p_T(j_2)>250~\mathrm{GeV}$ \\
			\bottomrule
		\end{tabular}
		\caption{Selection criteria used in the cut-based analysis.}
		\label{tab:cuts}
	\end{table}
	
	The cut-flow of the signal and background cross sections, expressed in fb, is shown in Table~\ref{tab:cutflow} for the three benchmark points. The results indicate that the chosen selections preserve a visible fraction of the signal but do not suppress the SM backgrounds sufficiently to reach discovery sensitivity. As shown in Table~\ref{tab:significance}, the resulting significances are $3.82\sigma$, $4.04\sigma$, and $1.88\sigma$ for BP1, BP2, and BP3, respectively. Thus, the $5\sigma$ discovery threshold is not reached even at the projected maximum CLIC luminosity of $\mathcal{L}=5~\mathrm{ab}^{-1}$. This limitation is driven by the small post-selection signal cross section, which is of order $10^{-2}~\mathrm{fb}$ after the baseline cuts.
	
	\begin{table}[t]
		\centering
		\begin{adjustbox}{width=\columnwidth}
			\begin{tabular}{lccc cccc}
				\toprule
				\multirow{2}{*}{Selection} & \multicolumn{3}{c}{Signal} & \multicolumn{4}{c}{Background} \\
				\cmidrule(lr){2-4}\cmidrule(lr){5-8}
				& BP1 & BP2 & BP3 & $ttbbj$ & $WWV$ & $ttj$ & $tWb$ \\
				\midrule
				Basic & 0.0180 & 0.0147 & 0.00648 & 0.0112 & 0.180 & 0.827 & 1.13 \\
				Cut 1 & 0.00753 & 0.00653 & 0.00279 & 0.00297 & $5.14\times10^{-5}$ & 0.0166 & 0.0126 \\
				Cut 2 & 0.00480 & 0.00492 & 0.00212 & 0.00090 & 0.00 & 0.00460 & 0.00174 \\
				Cut 3 & 0.00408 & 0.00434 & 0.00189 & 0.000618 & 0.00 & 0.00275 & 0.00111 \\
				\midrule
				Eff. [\%] & 23.8 & 21.4 & 25.6 & 5.49 & 0 & 0.332 & 0.098 \\
				\bottomrule
			\end{tabular}
		\end{adjustbox}
		\caption{Cut-flow of the signal and SM-background cross sections, in fb.}
		\label{tab:cutflow}
	\end{table}
	
	\begin{table}[t]
		\centering
		\begin{tabular}{lc}
			\toprule
			Benchmark point & Significance \\
			\midrule
			BP1 & 3.82 \\
			BP2 & 4.04 \\
			BP3 & 1.88 \\
			\bottomrule
		\end{tabular}
		\caption{Cut-based significances for the three benchmark points at $\sqrt{s}=3~\mathrm{TeV}$ and $\mathcal{L}=5~\mathrm{ab}^{-1}$.}
		\label{tab:significance}
	\end{table}
	
	\section{Machine-learning approach}
	\label{sec:ML}
	
	The cut-based analysis shows that none of the benchmark points reaches the $5\sigma$ discovery threshold, as summarized in Table~\ref{tab:significance}. We therefore investigate whether the sensitivity can be improved by exploiting multivariate information through machine-learning techniques. In particular, we consider a gradient-boosted decision tree implemented with \XGB~\cite{Chen:2016btl} and a multi-layer perceptron (MLP) neural network implemented with Keras~\cite{chollet2015keras}. These methods have been widely used in collider phenomenology~\cite{Wang:2020ips,Yang:2024aav,Benbrik:2026zgr,Hammad:2022lzo,Hammad:2025wst}, where they often provide stronger discrimination than rectangular cut-based selections by learning correlations among multiple kinematic observables.
	
	\begin{table}[t]
		\centering
		\begin{adjustbox}{width=\columnwidth}
			\begin{tabular}{ll}
				\toprule
				Variable & Description \\
				\midrule
				$p_T(\ell_1)$ & Transverse momentum of the leading lepton \\
				$p_T(b_1)$ & Transverse momentum of the leading $b$-jet \\
				$p_T(b_2)$ & Transverse momentum of the subleading $b$-jet \\
				$p_T(b_3)$ & Transverse momentum of the third-leading $b$-jet \\
				$N_b$ & Number of $b$-tagged jets \\
				$N_j$ & Number of reconstructed jets \\
				$\met/H_T$ & Ratio of missing transverse energy to $H_T$ \\
				$\Delta R(b_1,b_2)$ & Angular separation between the two leading $b$-jets \\
				$\Delta R(\ell_1,b_1)$ & Angular separation between the leading lepton and leading $b$-jet \\
				$\Delta R(b_1,j_1)$ & Angular separation between the leading $b$-jet and leading jet \\
				$\Delta\phi(\ell_1,\met)$ & Azimuthal separation between the leading lepton and $\met$ \\
				$\Delta\phi(b_1,\met)$ & Azimuthal separation between the leading $b$-jet and $\met$ \\
				$M_{\rm inv}(j_1,j_2)$ & Invariant mass of the two leading jets \\
				$M_{\rm inv}(b_1,j_1,j_2)$ & Invariant mass of the hadronic top candidate \\
				$M_{\rm inv}(b_1,b_2,j_1,j_2)$ & Invariant mass of the $b_1b_2j_1j_2$ system \\
				$\max(|\eta_j|)$ & Maximum absolute pseudorapidity among all jets \\
				\bottomrule
			\end{tabular}
		\end{adjustbox}
		\caption{Input observables used for the training and testing of the \XGB{} and MLP classifiers.}
		\label{tab:inputs}
	\end{table}
	
	For each benchmark point, we train a separate binary classifier using the variables listed in Table~\ref{tab:inputs}. The combined signal and background sample contains about $250{,}000$ simulated events. We split the sample into $80\%$ for training and $20\%$ for testing, using a stratified split to preserve the signal and background fractions. A validation subset of the training sample is used to choose the model, apply early stopping, and optimize the classifier-score threshold. The test sample is kept independent and is used only for the final performance and significance estimates.

	The \XGB{} classifier uses $10{,}000$ estimators, a learning rate of $0.01$, and a maximum tree depth of 3. The MLP is a fully connected network with three hidden layers containing 128, 64, and 32 neurons, respectively. Each hidden layer uses a rectified linear-unit activation. Batch normalization is applied after the first two hidden layers, and dropout rates of 0.30, 0.25, and 0.20 are used in the successive layers. The output layer contains one neuron with a sigmoid activation. The network minimizes the binary cross-entropy loss with the Adam optimizer and is trained for at most 300 epochs. Early stopping is based on the validation AUC, and the model with the best validation AUC is retained. This procedure keeps all choices that depend on the classifier output separate from the test sample and limits overfitting.
	
	\begin{figure*}[t]
		\centering
		\includegraphics[width=0.9\textwidth]{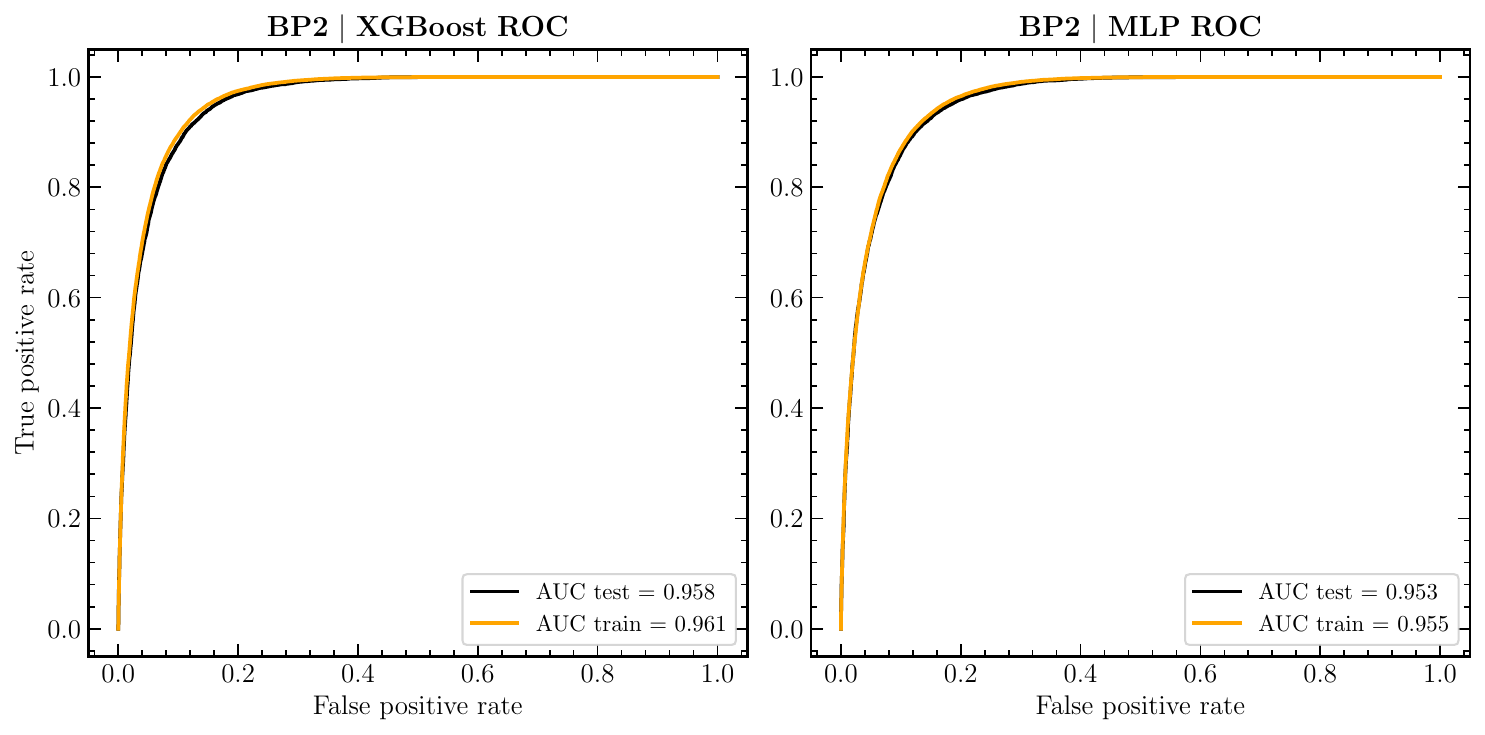}
		\caption{ROC curves for the BP2 benchmark point, comparing the \XGB{} and MLP classifiers. The true positive rate is shown as a function of the false positive rate for the training and test datasets. The corresponding AUC values are displayed in each panel.}
		\label{fig:roc}
	\end{figure*}
	
	The receiver operating characteristic (ROC) curves for the BP2 scenario are shown in Fig.~\ref{fig:roc} for both the \XGB{} and MLP classifiers. These curves display the true positive rate as a function of the false positive rate and provide a direct measure of the discriminating power between signal and background.
	
	For the \XGB{} classifier, excellent performance is obtained, with an area under the curve (AUC) of 0.961 for the training set and 0.958 for the test set. The close agreement between the training and test AUC values indicates good generalization and no significant overfitting. The MLP also achieves strong classification performance, with AUC values of 0.956 and 0.954 for the training and test datasets, respectively. Both models therefore exhibit high discriminating power, with \XGB{} showing a marginal but consistent advantage.
	
	\begin{figure*}[t]
		\centering
		\includegraphics[width=\textwidth]{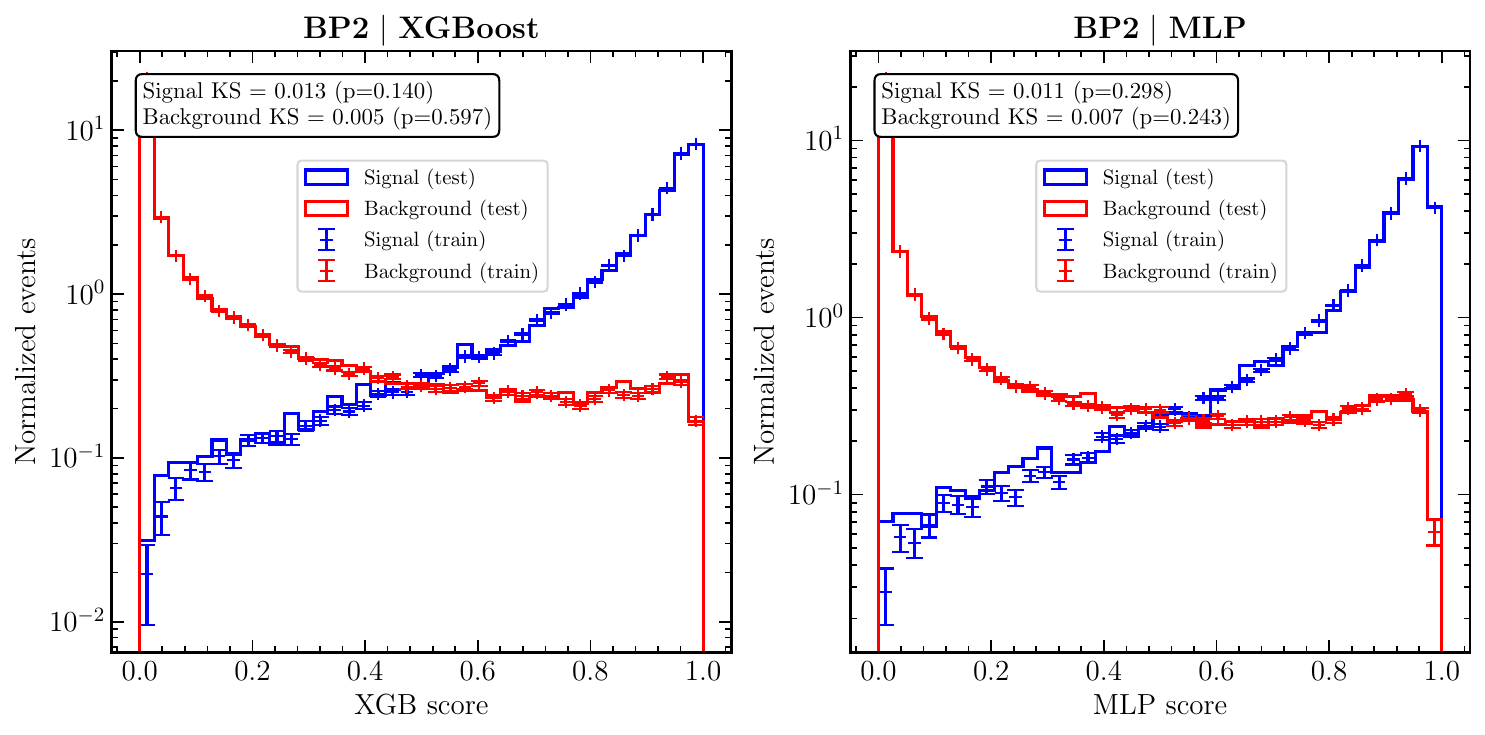}
		\caption{Normalized discriminator-score distributions for signal and background events in the BP2 scenario. The left panel shows the \XGB{} score, while the right panel shows the MLP score. Solid histograms and points with error bars correspond to the training and test samples, respectively.}
		\label{fig:score_distributions}
	\end{figure*}
	
	The normalized discriminator-score distributions for signal and background events in the BP2 benchmark point are shown in Fig.~\ref{fig:score_distributions}. The Kolmogorov-Smirnov (KS) test is used to quantify the agreement between the training and test distributions. This non-parametric test measures the maximum vertical distance between the empirical cumulative distribution functions of two samples,
	\begin{equation}
		D=\sup_x\left|F_1(x)-F_2(x)\right|,
	\end{equation}
	where $F_1(x)$ and $F_2(x)$ denote the empirical cumulative distribution functions. The associated $p$-value gives the probability, under the null hypothesis that the two samples are drawn from the same underlying distribution, of observing a KS statistic at least as large as the measured value. For the \XGB{} score, we obtain Signal KS $=0.013$ with $p=0.140$ and Background KS $=0.005$ with $p=0.597$. For the MLP score, we obtain Signal KS $=0.010$ with $p=0.420$ and Background KS $=0.006$ with $p=0.366$. These values confirm the excellent agreement between training and test distributions for both signal and background samples, indicating negligible overtraining in both classifiers.
	
	\begin{figure*}[t]
		\centering
		\includegraphics[width=\textwidth]{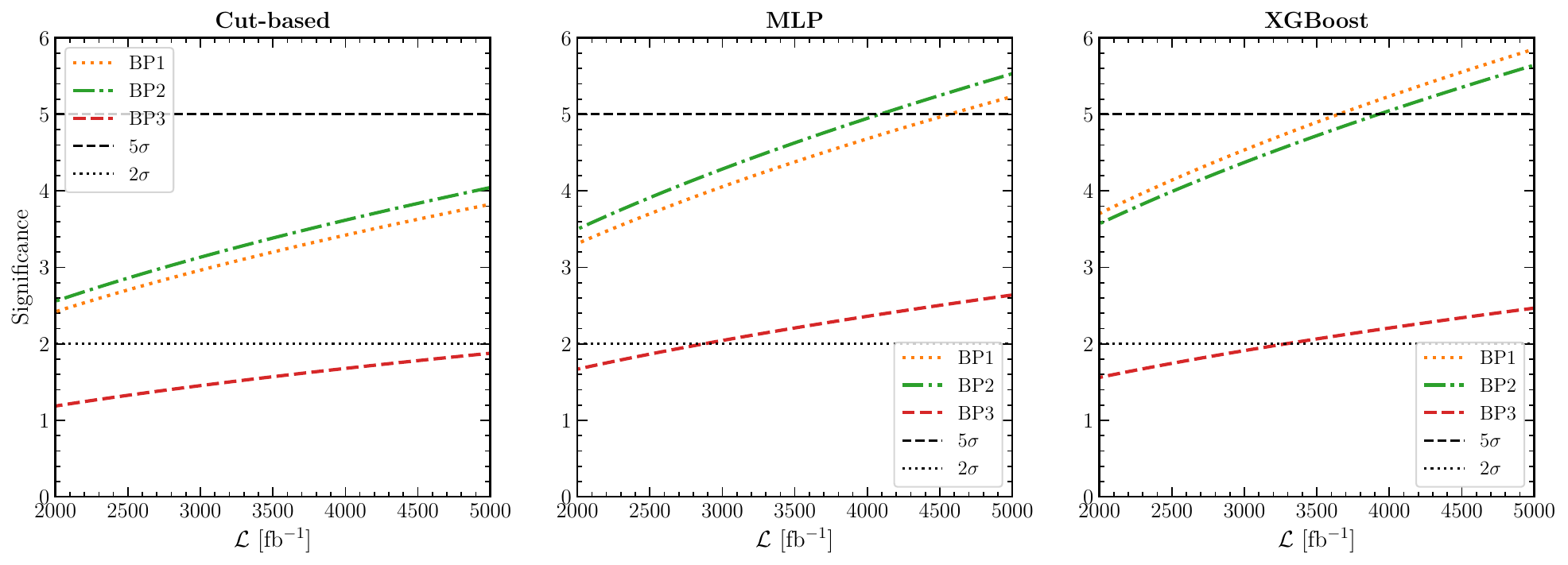}
		\caption{Expected signal significance as a function of the integrated luminosity $\mathcal{L}$ for the three benchmark points, comparing the cut-based analysis (left), the MLP classifier (middle), and the \XGB{} classifier (right). The curves are evaluated with Eq.~\eqref{equ:no_sys}, corresponding to $\delta=0$. The horizontal lines indicate the $5\sigma$ discovery and $2\sigma$ exclusion thresholds.}
		\label{fig:significance_vs_lumi}
	\end{figure*}
		
	\begin{table}[t]
		\centering
		\begin{adjustbox}{width=\columnwidth}
			\begin{tabular}{c ccc ccc}
				\toprule
				\multirow{2}{*}{$m_B$ [TeV]}
				& \multicolumn{3}{c}{MLP}
				& \multicolumn{3}{c}{\XGB} \\
				\cmidrule(lr){2-4}\cmidrule(lr){5-7}
				& $\delta=5\%$ & $\delta=10\%$ & $\delta=15\%$
				& $\delta=5\%$ & $\delta=10\%$ & $\delta=15\%$ \\
				\midrule
				1.0 & 3.48 & 3.42 & 3.33 & 5.29 & 5.24 & 5.16 \\
				1.1 & 5.68 & 5.59 & 5.46 & 5.82 & 5.69 & 5.50 \\
				1.2 & 5.67 & 5.54 & 5.34 & 6.66 & 6.48 & 6.21 \\
				1.3 & 6.27 & 6.06 & 5.76 & 6.47 & 6.26 & 5.95 \\
				1.4 & 6.12 & 5.93 & 5.65 & 6.24 & 6.03 & 5.73 \\
				1.5 & 5.80 & 5.64 & 5.39 & 5.82 & 5.62 & 5.33 \\
				1.6 & 4.56 & 4.45 & 4.30 & 5.17 & 5.02 & 4.80 \\
				1.7 & 4.39 & 4.34 & 4.27 & 3.87 & 3.74 & 3.56 \\
				1.8 & 3.65 & 3.59 & 3.49 & 3.11 & 3.01 & 2.87 \\
				1.9 & 2.85 & 2.81 & 2.76 & 2.45 & 2.37 & 2.26 \\
				2.0 & 1.76 & 1.73 & 1.67 & 2.01 & 1.96 & 1.88 \\
				\bottomrule
			\end{tabular}
		\end{adjustbox}
		\caption{Expected significance as a function of $m_B$ at $\mathcal{L}=5~\mathrm{ab}^{-1}$ for the MLP and \XGB{} analyses. The values are calculated from the signal and background yields in Table~\ref{tab:results} using Eq.~\eqref{equ:with_sys}, the classifier threshold is kept fixed while $\delta$ is varied.}
		\label{tab:significance_5000}
	\end{table}
	
	Table~\ref{tab:significance_5000} presents the signal significance as a function of $m_B$ for the MLP and \XGB{} analyses at $\mathcal{L}=5~\mathrm{ab}^{-1}$, including the impact of systematic uncertainties on the background yield. The significance is computed using the profile-likelihood-inspired expression~\cite{Cowan:2010js}
	\begin{align}
		\mathcal{Z} &= \Bigg\{2\Bigg[(s+b)\ln\!\left(\frac{(s+b)(1+\delta^2b)}{b+\delta^2b(s+b)}\right) \notag \\
		&\quad -\frac{1}{\delta^2}\ln\!\left(1+\delta^2\frac{s}{1+\delta^2b}\right)\Bigg]\Bigg\}^{1/2},
        \label{equ:with_sys}
	\end{align}
	where $s$ and $b$ denote the expected signal and background yields after the classifier-score requirement, and $\delta$ is the fractional uncertainty on the background normalization. The values in Table~\ref{tab:significance_5000} are recomputed directly from the yields in Table~\ref{tab:results}, using one fixed classifier threshold for each mass and classifier. In the limit $\delta\to0$, this expression reduces to
	\begin{align}
		\mathcal{Z}=\sqrt{2\left[(s+b)\ln\left(1+\frac{s}{b}\right)-s\right]}.
        \label{equ:no_sys}
	\end{align}
	
	The largest significances occur in the intermediate-mass region, where the increasing ${\rm BR}(B\to\phi b)$ and improved kinematic separation compensate for the falling production cross section. With a $15\%$ background-normalization uncertainty, the MLP remains above $5\sigma$ for the sampled masses $m_B=1.1$-$1.5~\mathrm{TeV}$, while \XGB{} remains above $5\sigma$ for $m_B=1.0$-$1.5~\mathrm{TeV}$. For a $10\%$ uncertainty, the \XGB{} reach extends to the sampled point at $m_B=1.6~\mathrm{TeV}$. These intervals refer to the discrete mass grid in Table~\ref{tab:significance_5000}; no interpolation between adjacent points is assumed.

	Fig.~\ref{fig:significance_vs_lumi} shows the luminosity dependence for the three benchmark points. The curves are obtained by scaling the signal and background yields with $\mathcal{L}$ and evaluating Eq.~\eqref{equ:no_sys}, i.e. with $\delta=0$. This choice isolates the relative performance of the cut-based, MLP, and \XGB{} analyses. The nonzero-systematic cases are presented separately in Table~\ref{tab:significance_5000}.

	The ordering of BP1 and BP2 changes slightly between the MLP and \XGB{} panels. To see whether this is a real classifier-dependent effect or simply a training fluctuation, we repeated each training 10 times with independent random seeds. In every run, the model and score threshold were chosen using the training and validation samples, while the final significance was evaluated on the independent test sample. At $5~\mathrm{ab}^{-1}$ we find
	\begin{align}
	Z_{\rm BP1}^{\rm MLP}&=5.44\pm0.34, & Z_{\rm BP2}^{\rm MLP}&=5.39\pm0.24,\\
	Z_{\rm BP1}^{\rm XGB}&=5.52\pm0.35, & Z_{\rm BP2}^{\rm XGB}&=5.22\pm0.37,
	\end{align}
	where the quoted uncertainties are the standard deviations over the 10 trainings. The difference between the mean BP1 and BP2 significances is only $0.05$ for the MLP and $0.30$ for \XGB{}, which is comparable to or smaller than the run-to-run spread. We therefore regard the ordering seen in the single-training curves of Fig.~\ref{fig:significance_vs_lumi} as a statistical fluctuation, not as evidence that either classifier systematically performs better for one benchmark point.
\section{Conclusions}
\label{sec:Conclusions}

We have studied the single production of a vector-like bottom quark at CLIC with $\sqrt{s}=3~\mathrm{TeV}$ in the $(T,B)$ doublet realization of the type-II 2HDM with VLQs. The cascade $e^+e^-\to B\bar b+\bar B b$, followed by $B\to\phi b$ and $\phi\to t\bar t$, probes a region that is not directly covered by searches assuming that the $B$ quark decays only to $Wt$, $Zb$, and $hb$.

For the three representative benchmark points, the cut-based analysis remains below $5\sigma$ at $5~\mathrm{ab}^{-1}$. Both multivariate classifiers improve the sensitivity considerably. Their training and test performances agree well in the ROC-AUC and Kolmogorov--Smirnov checks, and the repeated-training study shows that the small change in the BP1--BP2 ordering is consistent with normal run-to-run fluctuations.

On the discrete mass grid considered here, both classifiers retain discovery sensitivity up to $m_B=1.5~\mathrm{TeV}$ for a $15\%$ uncertainty on the background normalization. For a $10\%$ uncertainty, the \XGB{} analysis also remains above $5\sigma$ at the sampled point $m_B=1.6~\mathrm{TeV}$. The best sensitivity occurs at intermediate masses, where the larger nonstandard branching fraction and improved kinematic separation compensate for the falling production cross section. This makes heavy-Higgs decays of vector-like quarks a useful and complementary search channel at future high-energy lepton colliders.
	\vspace{-0.5cm}
\section*{Acknowledgments}
M. Boukidi acknowledges support from the Narodowe Centrum Nauki under OPUS Grant No.~2023/49/B/ST2/03862.

	\appendix
	\section{ML results and performance}
	\label{sec:appA}
	
	\begin{table}[h]
		\centering
		\begin{adjustbox}{max width=\columnwidth}
			\begin{tabular}{c c ccc ccc}
				\toprule
				\multirow{2}{*}{$m_B$ [TeV]} & \multirow{2}{*}{$\sigma$ [fb]} & \multicolumn{3}{c}{MLP} & \multicolumn{3}{c}{\XGB} \\
				\cmidrule(lr){3-5}\cmidrule(lr){6-8}
				& & AUC & $S$ & $B$ & AUC & $S$ & $B$ \\
				\midrule
				1.0 & $1.620\times10^{-2}$ & 0.9395 & 7.75 & 2.88 & 0.9400 & 8.60 & 0.86 \\
				1.1 & $1.895\times10^{-2}$ & 0.9488 & 11.96 & 1.74 & 0.9478 & 14.77 & 2.88 \\
				1.2 & $1.965\times10^{-2}$ & 0.9545 & 14.93 & 3.25 & 0.9531 & 19.00 & 3.57 \\
				1.3 & $1.873\times10^{-2}$ & 0.9573 & 19.73 & 4.87 & 0.9563 & 20.33 & 4.74 \\
				1.4 & $1.703\times10^{-2}$ & 0.9589 & 18.68 & 4.56 & 0.9579 & 19.84 & 5.01 \\
				1.5 & $1.483\times10^{-2}$ & 0.9620 & 16.80 & 4.13 & 0.9603 & 18.71 & 5.41 \\
				1.6 & $1.251\times10^{-2}$ & 0.9598 & 11.72 & 3.55 & 0.9594 & 15.26 & 4.67 \\
				1.7 & $1.037\times10^{-2}$ & 0.9587 & 7.55 & 1.23 & 0.9584 & 12.04 & 6.18 \\
				1.8 & $8.387\times10^{-3}$ & 0.9558 & 8.11 & 2.79 & 0.9571 & 9.36 & 6.25 \\
				1.9 & $6.703\times10^{-3}$ & 0.9572 & 5.37 & 2.11 & 0.9563 & 7.31 & 6.60 \\
				2.0 & $5.313\times10^{-3}$ & 0.9565 & 4.16 & 4.28 & 0.9571 & 5.36 & 5.42 \\
				\bottomrule
			\end{tabular}
		\end{adjustbox}
		\caption{Comparison of the MLP and \XGB{} classifiers for each signal benchmark mass: signal cross section, ROC-AUC, and expected signal and background yields at $\mathcal{L}=5~\mathrm{ab}^{-1}$.}
		\label{tab:results}
	\end{table}
	
	\bibliographystyle{apsrev4-2}
	\bibliography{main.bib}
	
\end{document}